\baselineskip 22pt
\def\vsk{\vskip 11pt}
\newcount\fnotecount
\fnotecount=0
\def\fnote#1{\global\advance\fnotecount by
1\footnote{${}^{\the\fnotecount}$}{#1}}
\newcount\EEK
\EEK=0
\def\eek{\global\advance\EEK by 1\eqno(\the\EEK )}

\def\title#1{\centerline{\bf #1}%
\vsk
\centerline{Adam D. Helfer}
\centerline{Department of Mathematics, University of Missouri, Columbia,
Missouri 65211, U.S.A.}%
}
\def\gtrsim{{\buildrel >\over\sim}}
\def\lesssim{{\buildrel <\over\sim}}

\def\d{{\rm d}}
\def\tint{{\int _{-\infty}^\infty{\widehat T}_{00}(t,0,0,0) b(t)\d t}}
\title{On Quantum Inequalities}
\vsk\vsk
\centerline{Abstract}
\vsk
Building on the ``quantum inequalities'' introduced by Ford, I
argue that the negative local energies encountered in quantum field
theory can only be observed by detectors with positive energies at least
as great in magnitude.  This means that operationally the total energy
density must be non--negative.  Like reasoning shows that, in a
similar operational sense, the dominant energy condition must hold:  any
timelike component of the
four--momentum density is positive.

\vsk\vsk\noindent PACS 03.65.Bz, 11.10.-z
\vsk\noindent Key words:  Energy density, negative energies, quantum
inequalities

\vfill\eject

\it Introduction.  \rm
A startling prediction of relativistic quantum field theory is that,
while the total energy of a system should be positive or zero, the energy
density, and hence the energy of a subsystem, can be negative~[1].
And indeed this possibility is present generically.
Even for a Klein--Gordon field on Minkowski space, for any
smooth compactly supported bump function $B$, the expectation values
$$\langle\int {\widehat T}_{00}B \, \d ^3x\rangle\, ,\eek$$
for ${\widehat T}_{00}$ the renormalized
Hamiltonian density, are unbounded below [2].  Thus
states with arbitrarily negative energy densities are always
available.  The set of states with this expectation value equal to
$-\infty$ is dense in the Hilbert space.

Although negative total energy densities have never been directly observed,
they have received extensive theoretical investigation because they
contravene a
basic tenet of classical physics.  Indeed, if there were no
restrictions on the negative energies achievable, there would be gross
macroscopic consequences:  an ordinary particle could absorb a negative
energy and become a tachyon; an isolated patch of negative energy would
give rise to a repulsive gravitational field; one could violate the
second law of thermodynamics by using negative energies to cool systems
without an increase of entropy~[3--6]; and the general--relativistic effects
might include traversable wormholes, ``warp drives'' and time
machines~[7--10].
Too, it is something of a puzzle why such states do not interfere with
the dynamics of the quantum fields:  why do not perturbations (which are
always present) send the field cascading through these negative--energy
states, with a corresponding release of positive--energy radiation?
It is a matter of common experience that such effects do not
occur, or at least not often, and therefore there must be some mechanism
restricting the production of negative energy densities, their
magnitudes, durations, or interactions with other matter.

Such a mechanism was first proposed by Ford, and has been
investigated by him and Roman~[3,5,11].
They argue that
(for free Bose fields in Minkowski space) the
negative energy $-\Delta E$ localizable in a time of order
$\Delta t$ should satisfy a \it quantum inequality \rm
$$\Delta E\Delta t\lesssim\hbar\, .\eek$$\xdef\frek{\the\EEK}%
These inequalities are powerful; they evidently limit the occurences of
negative energies considerably.  However, they do not as they stand seem
to be a full explanation.  For one thing, the argument for (\frek ) depends
on a certain ``coherence'' assumption, which, as we shall discuss below, is
not generally valid.\fnote{On the other hand, a weaker bound that Ford
establishes (our inequality (4), below),
which is suggestive of the averaged weak energy condition in
general relativity~[12], does not depend on this assumption.}
For another, it is not clear that simply
restricting the occurences of large negative energies to short times is
enough to rule out unphysical effects.  Indeed, explicit analyses of
attempts to violate the second law of thermodynamics indicate that while
the quantum inequalities play a key role, an equally important one is
played by limitations on the measuring devices.  (See refs. [3--6].
These also point out that identifying the
characteristic time $\Delta t$ which is relevant for a
particular physical problem may not be an easy matter.)

In this paper we shall re--examine the derivation of the quantum inequalities,
repair the ``coherence'' assumption,
and argue further that a device constructed to measure or capture a local
negative energy $-\Delta E$ must itself have energy at least \rm $\Delta E$.
We may say briefly that \it operationally the total energy must be
non--negative.  \rm   We
believe that this principle explains why there are no gross consequences of
negative energy density states, and underlies the more ad--hoc analyses which
have been offered to show that particular attempts to use negative energy
densities to violate the second law of thermodynamics fail.

Finally, we shall extend the reasoning to show that operationally the
dominant energy condition must hold, that is, the four--momentum density must
be future--pointing.  This gives a
new perspective on questions of how violations of the energy conditions
affect the usual general relativistic singularity and
positivity--of--energy theorems, as well as Hawking's process for the
evaporation of black holes, but these applications will be discussed
separately.  Apart from a few comments, the present paper is restricted to
special--relativistic quantum field theory.

Our metric signature is ${+}\,{-}\,{-}\,{-}$.

\it The Quantum Inequalities.  \rm
Ford and Roman have
given several derivations of the quantum inequalities,
but the elements which are relevant here are common to all.  Consider
the quantity
$$F=-\inf _{|\Phi\rangle}\langle\Phi |\int _{-\infty}^\infty
  {\widehat T}_{00}(t,0,0,0) \, b(t)\, \d t |\Phi\rangle\, ,\eek$$
(over normalized states $|\Phi\rangle$ in the Fock space of a
Klein--Gordon field in Minkowski space), where
$b(t)=t_0/[\pi (t^2+t_0^2)]$
is a ``sampling function'' with integral unity and characteristic width
$t_0$.\fnote{The
sampling function is peaked over an
interval of characteristic width $\sim t_0$, but not supported only there.
It is not possible to localize to
a sharply demarcated time interval, because the quantity
$\langle\Phi |{\widehat T}_{00}(t,0,0,0)|\Phi\rangle$ is a distribution,
and turns out to involve terms like $\delta '(t)$.
It is this which is behind
difficulties in identifying the correct $\Delta t$ for physical
applications.}  It is shown that
$$F\leq k\hbar c(ct_0)^{-4}\, ,\eek$$\xdef\Feek{\the\EEK}%
where it is known that the numerical
constant $k\leq 3/(32\pi ^2)$.
Up to this point the
argument is essentially mathematical.

The next step is physical.  If a device were to be constructed to
measure or trap this negative energy within an interval of length $t_0$,
then in order to function coherently the linear dimension of the device
must be no larger than $ct_0$.\fnote{Perhaps
one should use $ct_0/2$ here.  This would make the force of our later
arguments somewhat stronger.}  Thus the magnitude of the negative
energy within the device
$$\Delta E\leq F\cdot (4\pi /3)(ct_0)^3\leq (4\pi k/3)\hbar {t_0}^{-1}
  \leq \hbar /(8\pi t_0)\, .\eek$$\xdef\feek{\the\EEK}%
This hypothesis of coherent functioning deserves closer scrutiny.

The trouble here is that although it may be reasonable to think of an
experiment as a whole (including preparation at the start and collection
of data at the end) as ``coherent'' on a time scale $T_0$, the scales
$t_0$ of the components of the experiment may be much smaller.
For example,
suppose we had $N$ devices obeying (\feek ), so
capable of detecting or
trapping a negative energy
$-\varepsilon(4\pi k/3)\hbar {t_0}^{-1}$
in time $t_0$; here $\varepsilon <1$ is the
efficiency of the device.
These devices are arranged in
in an array in space, and in a common rest--frame.  Each carries a clock
which has been synchronized with (say) a master clock in the center of
the array.  At a preset time, each device operates.  Then, if the field
is in a suitable configuration, the total negative energy absorbed will be
$-N\varepsilon(4\pi k/3)\hbar {t_0}^{-1}$
and the interval will be $t_0$.  (Note that there is no
requirement that the devices be near one another, so they can be
separated far enough apart that locality considerations guarantee that
the quantum field can indeed be in a state which will produce such a
negative energy.  Also note that while it is true that construction of
the array of devices requires a different time scale than $t_0$, that
time scale is \it larger, \rm namely the time required
to synchronize the devices, greater than $\sim N^{1/3}t_0$.)
By choosing $N$ large enough, we can arrange for an arbitrarily large
negative energy to be trapped within a time $t_0$.
Thus even if we start from ``coherent'' devices, we can create others
which violate the quantum ineqality (\feek ).

We can repair this by taking into account the energy of the measuring
device.
A device which measures $\tint$ must involve some sort of clockwork
mechanism and transducer which function to weight
the contributions of ${\widehat T}_{ab}$ at different times by $b(t)$.
This clockwork must be able to resolve time increments of order $t_0$.
(Actually, in order to treat the function $b(t)$ as free from quantum
indeterminacy, the temporal resolution must be finer.)  Now let us
recall that a clock mechanism which is accurate to a time of order $\Delta
t$ must have mass $\gtrsim \hbar /(c^2 \Delta t)$ and so energy
$E_{\rm mech}\gtrsim\hbar /\Delta t$ [13].
For any one clock having energy $E_{\rm mech}$, then, controlling a
measuring device, the inequality (\feek ) applies with $t_0\sim\hbar
/E_{\rm mech}$.

This suggests that any device controlled by a clock of energy
$E_{\rm mech}$ can detect or trap negative energies $-\Delta E$ with
$-\Delta E+E_{\rm mech}\geq 0$ only.
It should be made clear that this argument is not a mathematical proof.
For one thing, the
quantities $\Delta t$ and $E_{\rm mech}$ are only defined as orders of
magnitude, and it is in this sense that $E_{\rm mech}\gtrsim\hbar /
\Delta t$ is known to hold.
For another, the quantum inequality (\feek ) as only been established
for one form of sampling function.  Nevertheless, the numerical factor
$1/8\pi$ in inequality (\feek ) is far enough below unity that
it strongly suggests $E_{\rm mech}\geq\Delta E$.
For the remainder of this paper, we shall assume this is the case.

With this assumption, notice that a collection of
measuring or trapping devices deployed
and set to function simultaneously (or, more generally, at spacelike
separations), as in the example above, will also have total energy in
excess of the negative energy it can detect or trap.\fnote{The case of
timelike separations requires a deeper analysis, taking into account
intermediate reductions of the state vector, and will be taken up
elsewhere.}

We may summarize our contention by saying that \it operationally, the
energy must be non--negative, \rm that is, the sum of the measured
energy and the energy of the measuring device must be non--negative.

\it Generality of the Results.  \rm
The analysis above is only for a Klein--Gordon field in Minkowski space,
but the result is of such a form that one is led to conjecture that it
holds generally.  We do not expect much difficulty in extending
it to higher--spin linear
fields.  Non--linear quantum field theories
in four dimensions are not well enough understood to be able to derive
rigorous estimates like Ford's $F\leq(3/32\pi ^2)\hbar c/(ct_0)^4$.
However, we offer a
few comments about them.

For electroweak theory (or any other weak coupling
renormalizable theory), at any one energy scale the theory is a
perturbation of free field theory, and our
arguments should apply to this effective theory.  One might hope to move
beyond this by applying renormalization--group techniques to test
whether Ford--type inequalities hold over a finite range of energies.

The situation for quantum chromodynamics is to some extent opposite.
There the low--energy theory is highly non--linear.  On the other hand,
quantum chromodynamics is expected to be asymptotically free, so its
short--distance behavior is that of a free theory, and one would expect
a Ford--type inequality to hold at least in the limit $t_0\to 0$.

For quantum fields in curved space--time, there is no
generally--accepted renormalization prescription for the stress--energy
operator.  Still, it is possible one could establish
a Ford--type inequality in
the limit $t_0\to 0$, since in this limit the renormalization
ambiguities become irrelevant and the asymptotic short--distance
behavior is that of free fields in Minkowski space.  On the other hand,
it is only the \it leading \rm term in the asymptotics which agrees with
the Minkowskian one; the subdominant terms in curved space--times are
more divergent than for Minkowski space~[2].  So this case requires
closer investigation.

\it The Dominant Energy Condition.  \rm
The treatment so far concerns the energy of a finite system, as measured
by an inertial observer.
The result localizes:  even if one tries to separate the clockwork used for
measuring $\tint$ from the world--line $(t,0,0,0)$, one must still transmit
timing signals to the vicinity of this world--line, and these signals must
resolve times $\lesssim t_0$, which means the quanta carrying the signals must
have energies $\gtrsim \hbar /t_0$.  Thus locally the total energy, of
the field plus the measuring device, must be non--negative.
We may call this the operational \it weak energy
condition: \rm  $T_{ab}^{\rm op}t^at^b\geq 0$ for all timelike vectors $t^a$.

It is possible to derive a stronger result, the operational \it dominant
energy condition:  \rm  $T_{ab}^{\rm op}t^au^b\geq  0$ for all
future--pointing vectors $t^a$ and $u^a$.  The changes needed to the treatment
above are as follows.

Let $t^a$ be a unit future--pointing vector along the observer's time axis.
Let $u^a$ be another unit future--pointing vector.  We shall be concerned with
a local measurement by the observer of a component $P_au^a$ of the momentum in
some region.

The following quantum inequality is easily
provable by the techniques of ref.~[11]:  Let
$$\Pi _a=\langle \int
_{-\infty}^\infty {\widehat T}_{a0}(t,0,0,0)b(t)\d t\rangle\, ;\eek$$
then
$\Pi _a u^a\geq -(3/32\pi ^2)t\cdot u\hbar c/(ct_0)^4$
for any future--pointing vector
$u^a$, so
$P_au^a\geq -(1/8\pi )t_au^a\hbar /t_0$.  Equivalently,
$$P_a= -(1/8\pi )(\hbar /t_0)t_a+\hbox{  a future--pointing vector}\, .\eek$$

Now consider a clock which may be boosted relative to $t^a$.  If the clock is
required to have resolution $\Delta t$ in the $t^a$--frame, then its
resolution in its own frame must be $\Delta t/\gamma$, with $\gamma$ the
usual Lorentz factor.  Its mass must be
$$m\gtrsim\hbar\gamma /(c^2\Delta t)\, .\eek$$\xdef\mineq{\the\EEK}%
Let the clock's four--momentum $P_a$ be $(E,p)$ in the
$t^a$--frame, so $E=mc^2\gamma$ and $p=mc\beta\gamma$.  Then
$$E^2-mc^2E/\gamma -p^2c^2=0\eek$$
from which
$$\bigl( E-mc^2/(2\gamma)\bigr) ^2 -p^2c^2=m^2c^4/(4\gamma ^2)\, .\eek$$
This means
$$P^{\rm clock}_a=mc^2/(2\gamma )t_a +\pi _a\, ,\eek$$
where $mc^2/(2\gamma )\gtrsim \hbar /(2\Delta t)$ and $\pi _a$ is timelike
future--pointing with $\pi _a\pi ^a
=(mc^2/2\gamma )^2\gtrsim (\hbar /2\Delta t)^2$.

Combining the results of the two previous paragraphs, we see that for any
future--pointing vector $u^a$, the sum of the expectation value $P_au^a$ of
the $u^a$--component of the momentum and the corresponding component of the
momentum of the clock which controls the sampling satisfies
$$\bigl( P_a+P^{\rm clock}_a\bigr) u^a
    \geq mc^2/\gamma -(1/8\pi )\hbar /t_0\, ,\eek$$
which we expect to be positive by (\mineq ).

A word about the interpretation of this is in order.  Here $P_a$ is
the expectation of ${\widehat T}_{ab}t^b$ smeared over a volume in
Minkowski space.  The components of this smeared operator
do not generally
commute (one cannot simultaneously measure the components of the
four--momentum in a finite box, because of edge effects).  Thus it
perhaps too strong to say that the four--momentum is operationally
future--pointing, since the four--momentum of the field within a finite
box cannot, strictly speaking, be measured.  What we have shown is that
for any future--pointing vector $u^a$, the operator $u^aP_a^{\rm op}$
is non--negative,
where $P_a^{\rm op}$ is the sum, of the clock's four--momentum and the
four--momentum operator for the field in a box.

\it Conclusions. \rm
The operational dominant energy condition immediately resolves two of the
negative--energy pathologies listed in the introduction.
It precludes the conversion of ordinary particles to tachyons.
And
at the level of Newtonian gravity, it forces gravitational fields to be
attractive in the sense that ${\bf\nabla}\cdot{\bf g}\leq 0 $, for ${\bf g}$
the gravitational acceleration field, since a measurement of ${\bf\nabla}\cdot
{\bf g}$ is a measurement of the energy density.

The remaining issues require more extensive discussion than can be given
here.  We can only comment briefly
that Grove~[6], in his resolution of the second--law problems
raised by Ford~[3] and Davies~[4] in effect establishes a special case of the
operational positivity of energy.
We hope to discuss this, and the
question of why perturbations do not cause quantum systems to
decay into states with patches of negative energy together
with positive--energy radiation, in a future publication.

Finally, the applications of this to theory of quantum fields in curved
space--time, particularly to the question of what the back--reaction of
the fields on the space--time geometry is, will be discussed elsewhere~[14].

\vsk\vsk
\noindent\bf Acknowledgement\rm
\vsk
\noindent I am grateul to members of the University of
Missouri--Columbia
Relativity
Group, particularly Bahram Mashhoon, for useful comments.

\vsk\vsk
\noindent\hrule width 3in
\vsk
\centerline{\bf References}
\frenchspacing

\parshape=0
\global\parindent=0pt
\everypar{\hangafter=1\hangindent=3pc}

[1] H. Epstein, V. Glaser and A. Jaffe, Nuovo Cimento \bf 36 \rm (1965)
1016--22.

[2] A. D. Helfer, Class. Quant. Grav. \bf 13 \rm (1996) L129--L124.

[3] L. H. Ford, Proc. R. Soc. Lond. \bf A364 \rm (1978) 227.

[4] P. C. W. Davies, Pys. Lett. \bf 113B \rm (1982) 215--218.

[5] L. H. Ford, Phys. Rev. \bf 43D \rm (1991) 3972--3978.

[6] P. G. Grove, Class. Quant. Grav. \bf 5 \rm (1988) 1381--1391.

[7] M. Morris and K. Thorne, Am J. Phys. \bf 56 \rm (1988) 395.

[8] M. Morris, K. Thorne and Y. Yurtsever, Phys. Rev. Lett. \bf 61 \rm
(1988) 1446.

[9] M. Alcubierre, Class. Quant. Grav. \bf 11 \rm (1994) L73.

[10] A. Everett, Phys. Rev. \bf D53 \rm (1996) 7365.

[11] L. H. Ford and T. A. Roman, \it Restrictions on Negative Energy
Density in Flat Spacetime, \rm preprint, gr--qc 9607003.

[12] E. Flanagan and R. M. Wald, \it Does backreaction enforce the
averaged null energy condition in semiclassical gravity?\rm , preprint,
gr-qc 9602052.

[13] H. Salecker and E. P. Wigner, Phys. Rev. \bf 109 \rm (1958)
571--577.

[14] A. D. Helfer, in preparation.

\bye